\documentstyle[eqsecnum,aps,12pt,prb]{revtex}
\input psfig.sty

\begin{document} 
 
\title{Coherent Mechanism of Robust Population Inversion}
\author{J. Vala, and  R. Kosloff} 
\address{Fritz Haber Research Center for Molecular Dynamics and
Department of Physical Chemistry,\\ 
Hebrew University, Jerusalem, 91904,
Israel}

\tighten
\maketitle 
\vspace{2.cm}

\begin{abstract} 
A coherent mechanism of robust population inversion in atomic and 
molecular systems by a chirped field is presented. It is demonstrated
that a field of sufficiently high chirp rate imposes a certain relative phase
between a ground and excited state wavefunction of a two-level system.
The value of the relative phase angle is thus restricted
to be negative and close to 0 or $-\pi$ for positive and negative
chirp, respectively. This explains the unidirectionality of 
the population transfer from the ground to the excited state.
In a molecular system composed of a ground and excited potential energy surface
the symmetry between the action of a pulse with 
a large positive and negative chirp  is broken.
The same framwork of the coherent mechanism can explain
the symmetry breaking and the population inversion due to a positive
chirped field. 

\end{abstract}

\newpage

Radiation induced transfer, of ground state population, 
to a designated excited state is an extremely useful quantum manipulation. 
Such a manipulation sets the stage for
subsequent clean experiments which are free from disturbances from ground state
population. In coherent control as well as in quantum computing
a complete population transfer can be used to prepare an almost pure initial state
which is a prerequisite for many manipulations.

The direct approach to population inversion by applying a $\pi$ pulse is 
plagued by the inability to  precisely control the intensity and time duration
of the excitation pulse. Moreover inhomogeneity in the sample, in particular the
orientation of the transition dipoles related to the impinging field
further hinder the direct approach. For a two-level-system (TLS) the solution
to the inversion problem is known as adiabatic following \cite{allen87,eberly99,band94}. 
Using a sufficiently chirped field so that the change in instantaneous frequency
is small relative to the Rabi frequency, adiabatic conditions prevail
resulting in a unidirectional population transfer.

For a molecular system composed of two potential surfaces (TPS) the problem 
of robust complete population transfer from the ground electronic state
to the excited one becomes more difficult. Cao, Bardeen and Wilson\cite{cao98,cao00}
suggested a population inversion scheme also based on a chirped excitation
pulse. Their explanation was given in the form of a wavepacket picture. 
A chirp leads to prolongation of the pulse duration while conserving the
total spectral band-width. In the case of linear chirp,
a new, instantaneous band-width can be identified as the reciprocal
value of the total pulse duration. Population transfer
between two molecular potentials then has two aspects. The total
band-width of the pulse addresses both potentials at a certain range 
of the molecular internuclear distance where available
Bohr frequencies correspond to the pulse spectrum. However the population 
transfer itself takes place through narrower coupling window defined by the
instantaneous bandwidth. In the case of positive chirp, the ground
state population is sequentially promoted to the excited state 
through the instantaneous window which moves from lower to higher 
interpotential energy difference. However, the excited state wavepacket
moves in the opposite direction due to the potential gradient.
Thus a new portion of the wavepacket, just being transferred, does not 
interfere with the old one preventing the stimulated emission from
happening. The result is the population inversion between two molecular 
electronic Born-Oppenheimer potentials.

In the present letter we present a unified viewpoint based 
on a simple coherent control analysis which can explain the mechanism
of robust complete population transfer in both the two-level system (TLS) as
well as for the two potential surface (TPS) scenario. 
The basic idea is to employ a control local in time which monotonically directs
the system to its final objective.

A robust manipulation is obtained if a unidirectional approach toward the
target of control is maintained. In the present case we want
to control the total change in population $\frac{d N_g}{dt}$. 
Where $N_g$ is either the ground state
population in the TLS or the expectation of the projection 
on the ground electronic surface i.e. 
$ N_g = \langle {\bf \hat P_g } \rangle ~=~ \int \vert \psi_g(r) \vert^2 dr $. 
In both cases the Heisenberg equation of motion for the change of the ground state
population induced by an electromagnetic field becomes \cite{kosloff92,rice00}:
\begin{equation}
\frac{dN_g}{dt} ~=~ \frac{2}{\hbar}  \vert \langle \psi_e \vert \hat{\mu} \vert \psi_g \rangle \vert \vert E(t) \vert sin(\phi_\mu + \phi_E)
\label{eom_pop}
\end{equation}
where $\langle \psi_e \vert \hat{\mu} \vert \psi_g \rangle$ is the transition
dipole moment, E(t) is the electromagnetic field and $\phi_\mu$ and $\phi_E$
are phase angles of the transition dipole and the field, respectively.
$\psi_i$, {\it i=g,e} is a ground and excited state wavefunction, respectively. 

Applying Eq. (\ref{eom_pop}) to the unidirectional population transfer 
$\psi_g \rightarrow \psi_e$ implies $\frac{dN_g}{dt} \le 0$,
which is controlled by the sum of phase angles $\phi_\mu$ and $\phi_E$.
The phase of transition dipole $\langle \psi_e \vert \hat{\mu} \vert \psi_g \rangle$
is assembled during the excitation process and is therefore 
a function of the history of the amplitude and phase of the 
excitation field. If initially all the population
resides on the ground state its initial phase has no relevance
since it does not alter the phase of the transition dipole.

A chirped electromagnetic field of Gaussian envelope in frequency representation 
has the following form \cite{cao98}

\begin{equation}
\tilde E(\omega ) = \tilde E(\omega_0) \exp[ -\frac{(\omega-\omega_0)^2}{2\Gamma^2} - i\chi' \frac{(\omega-\omega_0)^2}{2}]
~~~,
\end{equation}
where $\omega_0$ is the transform-limited carrier frequency of the field,
$\Gamma$ is the spectral bandwidth of the pulse and $\chi'$ is the chirp
rate in energy representation given by $dt/d\omega$. 
The chirp rate term causes a phase shift 
of each spectral component of the field proportional to its 'distance'
from the carrier frequency. The field in time representation is given
by its Fourier transform

\begin{equation}
E(t) =  E_0 exp[ -\frac{t^2}{2\tau^2} - i\omega_0 t - i\chi \frac{t^2}{2} + i\phi_E]~~~,
\end{equation}
where $\chi$ is the linear chirp rate in the time representation $d\omega/dt$. 
Chirp results  in prolongation of the pulse in time domain
reducing local field intensity to conserve the total pulse energy
$\tau^2 = 1/\Gamma ^2 + \Gamma ^2 \chi'^2$.
The chirp rate in time and frequency representations are then related
by the formula $\chi = \chi ' . \Gamma^2/\tau^2 $.

Without loosing generality $\phi_E $ can be set to zero
because a constant phase of the field maps onto the phase of the transition
dipole moment.
In this case the direction of the population transfer will be determined by the 
induced instantaneous phase of the transition dipole
$\langle \psi_e \vert \hat{\mu} \vert \psi_g \rangle = 
\vert \langle \psi_e \vert \hat{\mu} \vert \psi_g \rangle \vert e^{i \phi_\mu}$.
A phase angle $ -\pi  < \phi_{\mu} < 0 $ throughout the process will guarantee
a monotonic and robust population transfer. 

To verify this insight on the mechanism, a numerical scheme to solve
the time dependent Sch\"odinger equation (TDSE) 

\begin{eqnarray}
i \hbar {\partial \over{\partial t}}
\left( 
\begin{array}{c}
\psi_e \\
\psi_g
\end{array}
\right) ~~=~~ 
\left(
\begin{array}{cc}
\bf {\hat  H_e}  & - \mu E(t)\\
- \mu E(t)^* & \bf {\hat H_g}
\end{array} 
\right)
\left(
\begin{array}{c}
\psi_e \\
\psi_g
\end{array}
\right)
\label{eq:tdse}
\end{eqnarray}
is employed for both the TLS and TPS.
An initial state is propagated in time using a Chebychev polynomial
expansion of the evolution operator \cite{kosloff94}. 
The propagation was realized in discrete steps with a time increment 
shorter by two-orders of magnitude than the pulse duration. 
For a TPS, we used the Fourier grid representation of 
the wavefunction  and the quantum operators \cite{kosloff96}.
Typical computation parameters, summarized in Tab. \ref{parameters}, 
were chosen by the  criteria  of a correct representation 
of the wavepacket in coordinate as well as in momentum spaces.

\newpage
\begin{table}[h]
\caption{The typical values of the propagation parameters. The grid
parameters are related only to a TPS.}
\begin{center}
\begin{tabular}{||c|c|c||}\hline
Parameters & Typical value & Units \\
\hline
\hline
Time step & 4$\pi$/10 & a.u. \\
\hline
Number of time steps & 200 & \\
\hline
Pulse duration & $\sqrt{\pi/2}$ & a.u.\\
\hline
Typical maximal Rabi frequency & 1 & a.u.\\
\hline
Width of the initial wavepacket & 0.1 & a.u.\\
\hline
Initial position of the wavepacket & 0 & a.u.\\
\hline
Grid spacing & 0.05 & a.u.\\
\hline
Number of grid points & 256 & \\
\hline
\end{tabular}
\label{parameters}
\end{center}
\end{table}

\begin{figure}
\psfig{figure=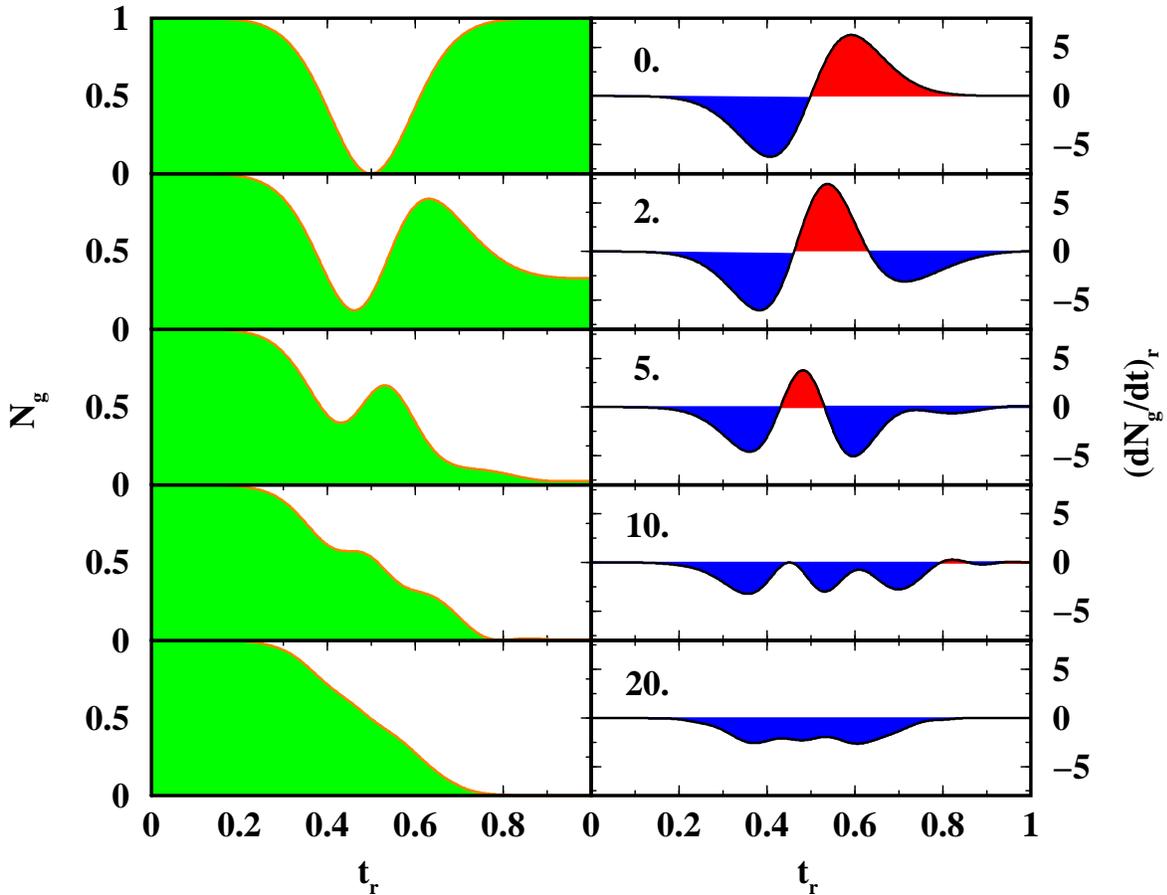,width=.9\textwidth}
\caption{The evolution of the ground state population $N_g(t)$ (left)
and the imaginary part of the transition dipole moment multiplied by 
the field  amplitude (right). The normalized
time is defined as $t_r = t / (6 \tau_{\chi'=0}.f)$ and 
$(dN_g/dt)_r = (6 \tau_{\chi'=0}.f) dN_g/dt$ is the rate of change in the normalized 
time units. 
$\tau_{\chi'=0}$ is the pulse duration for a transform-limited pulse ($\chi'=0$) 
and $f$ is the ratio of pulse duration between the chirped and  unchirped cases.
The numbers indicate the value of the chirp rate.} 
\label{tls_chirp}
\end{figure}

The reference TLS case is obtained by an  on-resonant transform-limited
Gaussian pulse, with integrated intensity
which causes a complete $2 \pi$ cycling from the ground state to
the excited state and back. The top panel of Fig. \ref{tls_chirp} 
shows the total population on the ground state $N_g$ as well as its
time derivative $\frac{d N_g}{dt }$, both obtained independently
from quantum wavefunctions propagated in time.
The phase or the imaginary part the transition dipole moment 
gains first negative values making $ dN_g/dt < 0 $. 
Once all the population is transfered to
the excited state, the imaginary part of the transition dipole moment
changes sign and redirect the population flow back to the ground state.

The lower panels show the same process upon increasing the chirp rate.
It is evident that the symmetry between excitation and deexcitation
is broken leading eventually with sufficient chirp rate,
to a monotonic population transfer. The imaginary
part of the transition dipole moment is restricted to
negative values in that case.

\begin{figure}
\psfig{figure=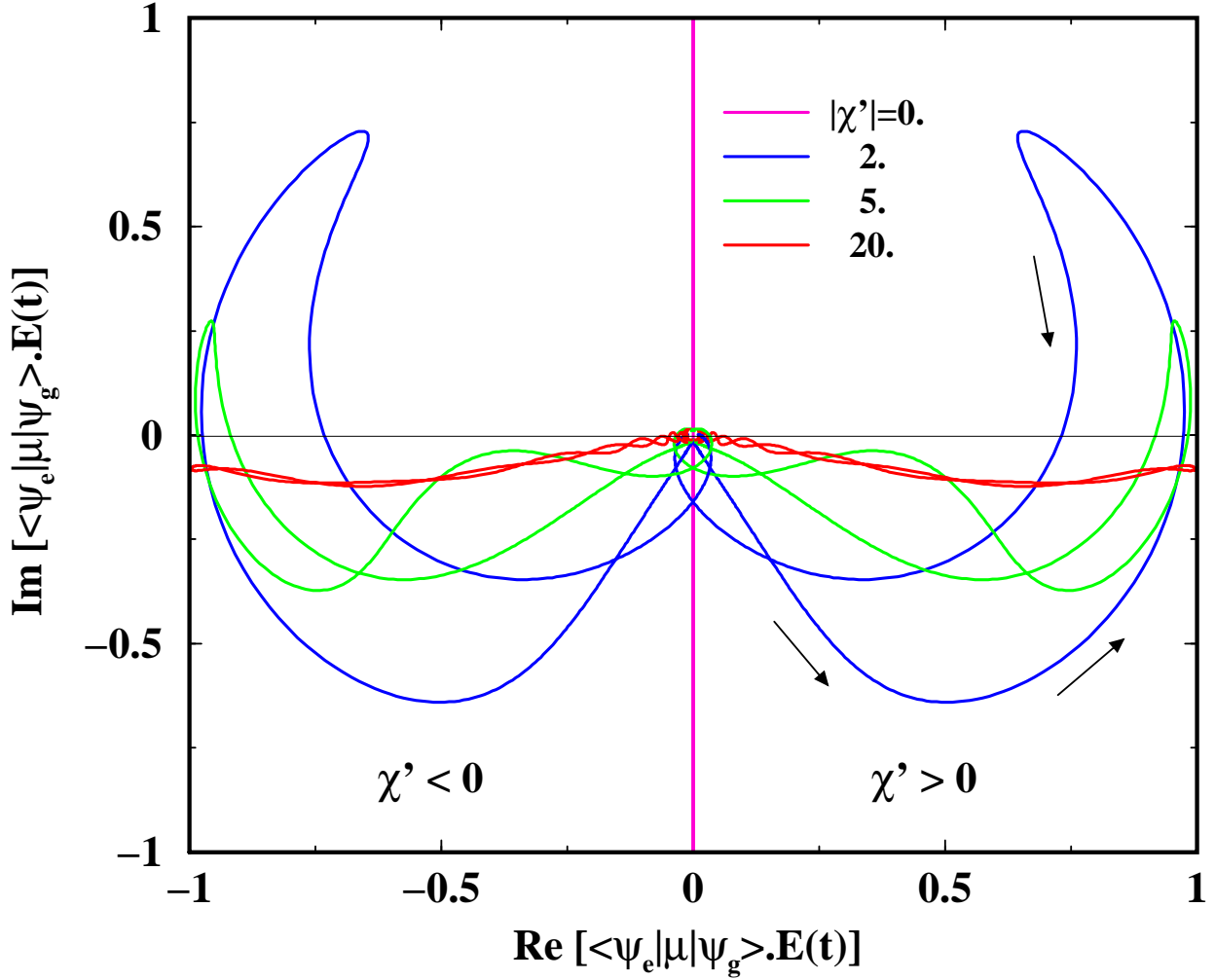,width=1.\textwidth}
\caption{Trajectories of the transition dipole moment, renormalized 
by its maximal amplitude, for excitation by the transform limited 
(on the imaginary axis) and
chirped pulsed field of positive and negative chirp rates.}
\label{traj}
\end{figure}

Fig. \ref{traj} displays the transition dipole trajectories 
during the excitation process.
For the  on-resonant transform-limited
pulse, the trajectory lies on the imaginary axis and the corresponding
relative phase angle switches between $-\pi/2$ and $\pi/2$.
With increasing chirp rate the trajectories also obtain a real component 
positive for the positive chirp and {\it vice versa}.
For the chirped field case more time is spent
in the negative imaginary part of the complex plane. 
With sufficient chirp the whole trajectory is maintained in the
negative imaginary quadrants.
The perfect symmetry of the trajectories with respect to pulses
with positive and negative chirp is obvious for this TLS case.

Chirped field excitation in a molecular systems is more involved due
to the inherent dynamics of nuclear wavepackets on the ground and excited
electronic potentials. A two-potential model is studied
consisting of a flat ground state and  a linearly 
decreasing excited state potential.
The field is assumed to be 
sufficiently broad band to address all the initial 
nuclear configurations on the ground electronic state. 
The intensity of the field is sufficient to transfer all the population.

Positive chirp results in complete population inversion. The instantaneous
coupling window given by the instantaneous frequency of the chirped field
moves in the direction from lower to higher Bohr frequency of the system.
The excited wavepacket moves in the opposite direction due to
the excited potential gradient\cite{cao98,cao00}.
The evolution of the imaginary part of the transition dipole 
is  restricted to negative values locally in the instantaneous
coupling window and thus the population  transfer
is unidirectional from the ground electronic potential.

\begin{figure}
\psfig{figure=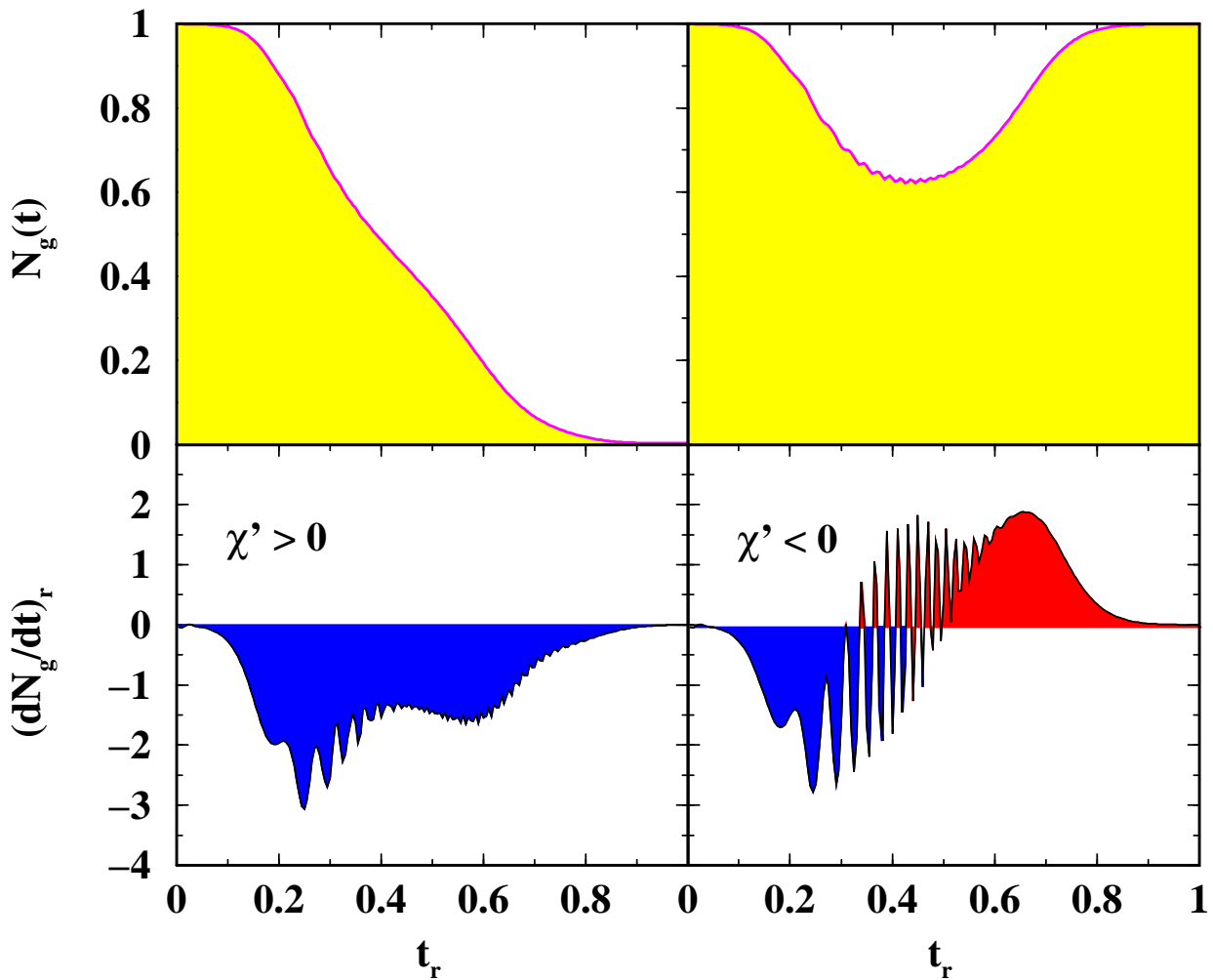,width=1.\textwidth}
\caption{Chirped field population transfer for $\chi' = 20$ in a two-potential 
system $V_g(r) = 0.$ and $V_e(r) = -2 . r$.
The positive chirp rate results in robust population inversion (invariant
with variation of field and system parameters in a certain extent). 
On the other hand, the negative chirp rate leads to the break-down of 
robust population transfer efficiency and can give any possible result
in dependence on the parameters (here the zero transfer is
apparently accidental).
The relative quantities are defined in 
Fig. \ref{tls_chirp}, the intensity is stronger by the factor of five.}
\label{sym_break}
\end{figure}

Negative chirp leads to a different result. The coupling window now moves
in the same direction as the excited state population. Due to this
evolution the excited wave- packet gains an extra phase which breaks
the subtle phase relations between the nuclear populations on the
electronic potentials coupled by the field. The corresponding
evolution of the imaginary part of the transition dipole moment
is thus not restricted to negative values and the population transfer
is not unidirectional as can be seen in  Fig. \ref{sym_break}.

To understand the physical origin of this behavior, the TDSE 
for the excited state wavefunction, which is initially of zero amplitude, 
is solved formally,

\begin{equation}
\vert \psi_e (t) \rangle ~~=~~ 
i \int_{0}^{t} 
d \tau  ~e^{ -\frac{i}{\hbar} {\bf \hat H_e } (t - \tau ) } 
\hat \mu E( \tau )  \vert \psi_g (\tau )\rangle
\label{eq:solution}
\end{equation}

The result is then used to reformulate the rate of the ground state population 
flow given in Eq. \ref{eom_pop},

\begin{equation}
\frac{dN_g}{dt} ~=~ - \frac{2}{\hbar} 
Im \left[ i \int_{0}^{t} d \tau  
\langle \psi_g (t) \vert  \hat{\mu} 
e^{ -\frac{i}{\hbar} {\bf \hat H_e } (t - \tau ) } \hat{\mu} 
\vert \psi_g (\tau )\rangle E^*(t) E(\tau) \right]
\label{pop_tdse}
\end{equation}

For a TLS coupled by an on-resonant field, the $\vert \psi_g (t) \rangle$
is known to be $ \cos(\vert \hat{\mu} E(t) \vert t) \vert \psi_g (0)\rangle$
\cite{ashkenazi97}. The field term in the Eq. \ref{pop_tdse}, proportional
to $e^{i \omega (t - \tau)}$, cancels the excited state free evolution term 
and the integration leads to the following expression

\begin{equation}
\frac{dN_g}{dt} ~=~ - \frac{2}{\hbar} 
\langle \psi_g (0) \vert \hat{\mu}^2  \vert \psi_g (0)\rangle E_0 \sin(2 \Omega t)
\label{pop_tdse_onr}
\end{equation}
where we used a square pulse $E_0 e^{-i \omega (t)}$ for the sake of 
simplicity. This solution shows that the alternation of the relative phase
angle between the ground and excited state TLS wavefunctions 
due to the on-resonant Rabi cycling is strictly $\mp \pi/2$ in accord
with our numerical results, see Fig. \ref{traj}. 

Equivalently,
the field exerts always the same relative phase between the levels
from and to which the population is being transferred, and that is $+ \pi/2$.
Since the relative phase concept does not have any meaning when only one of
the levels of a TLS is populated, the turning points of the on-resonant
Rabi cycling are singular from the point of view of the relative phase.

In the case of a transform-limited off-resonant population transfer, 
an additional
oscillatory term $e^{-i \Delta (t - \tau)}$
appears inside the integral in Eq. \ref{pop_tdse} due to
the difference  $\Delta$ between the system Bohr 
frequency ($\hbar \bf {\hat H_e}$) and
the frequency of the field. This term leads to an extra rotation of  
the relative phase in one direction and results in faster and less efficient
population transfer. 

A linear chirping of the field leads to a variation of the detuning with
time which changes its sign and hence inverts the relative phase rotation
in the middle of the process.
As the detuning approaches the resonance point, the extra phase rotation
slows down and eventually changes its direction. 
As a result, the relative phase
between both levels is pushed back to values in the negative
imaginary half of the complex plane and the population transfer stays
unidirectional.

An extra feature in the two potential model (TPS) compared to 
the two level system (TLS) is the wavepacket dynamics on both 
the ground and excited  electronic potentials.
This dynamics induces an additional relative phase shift which
then modifies the total relative phase relation between the ground and
excited state wavefunction and hence the population transfer.
This is particularly important when the timescale of the dynamics 
is comparable or higher than the timescale of the coupling
window motion given by the chirp rate, and both have the same orientation,
{\it i.e.} the case of the negative chirp rate. The process then
does not result in robust population inversion and can lead to
any result according to the field and molecular parameters.

If the dynamics are slow compared to the pulse duration this 
additional phase shift can be ommitted. This is the case 
in photoassociation of ultra cold atoms where the dynamics
is slowed down \cite{vala01}. Under these conditions
the TPS is continuously approaching the TLS limit. Alternatively,
the same effect can be achieved in ultrafast spectroscopy
where  the pulse duration can be made shorter than 
the dynamics \cite{rice00,ashkenazi97}.

\section*{Acknowledgments}

This research was supported by  the  Israel Science Foundation 
administered by the Israel Academy of Science.
The Fritz Haber Research Center is supported
by the Minerva Gesellschaft f\"{u}r die Forschung, GmbH M\"{u}nchen, FRG.

\end{document}